\documentclass[aps,showpacs,twocolumn]{revtex4}
\usepackage{epsfig}
\usepackage{graphicx,color,dcolumn}
\usepackage{epstopdf}
\usepackage{amsmath,amssymb}
\usepackage{multirow}
\usepackage{changes}
\usepackage{booktabs}
\usepackage{threeparttable}

\renewcommand{\arraystretch}{1.5} 
\begin{document}

\title{$X(3872)$ in an unquenched quark model}

\author{Yue Tan}\email{161002010@stu.njnu.edu.cn}
\author{Jialun Ping}\email{jlping@njnu.edu.cn(Corresponding author)}

\affiliation{Department of Physics and Jiangsu Key Laboratory for Numerical
Simulation of Large Scale Complex Systems, Nanjing Normal University, Nanjing 210023, P. R. China}

\begin{abstract}
  In this paper, we calculate mass and probability fractions of meson-meson components of $X(3872)$ in an unquenched
  quark model. Different from most of other unquenched quark models, the quark pairs creation operator from $^3P_0$ is
  modified by considering effects of the created quark pair energy and separation between the created quark pair and the
  valence quark-pair. In the calculation, all the wavefunctions of mesons and the relative motion between two mesons are
  obtained by solving the corresponding Schr\"{o}dinger equation with the help of gaussian expansion method.
  The multi-channel coupling of quark-antiquark state with possible meson-meson states are performed.
  The results show that the $X(3872)$ can be described as a mixing state of dominant charmonium state (70\%) and
  meson-meson components (30\%).
\end{abstract}

\pacs{14.40.Be,12.39.Pn,24.10.Eq}

\maketitle

\section{Introduction} \label{introduction}
The nonrelativistic quark model has been successfully describing the properties of heavy mesons. However,
since the well-known exotic state, $X(3872)$, was discovered, more and more exotic particles which cannot be
fitted well into meson spectrum, are reported by experimental collaborations. These exotic states put
forward great challenge on quark model.

In 2003, the Belle collaboration first found $X(3872)$ in the $B$ meson decay~\cite{01Choi:2003}. Subsequently,
CDF~\cite{02Acosta:2004} and D0~\cite{03Abazov:2004} collaborations confirmed this state in $p\bar{p}$ collision,
and BABAR also found $X(3872)$ in the $B$ meson decay~\cite{04Aubert:2005}. Different from the ordinary hadrons,
the state $X(3872)$ has some strange properties, its mass is very close to the threshold of $D\bar{D}^*$,
and decay width is very narrow, less than 1.2 MeV. So the dispute on the nature of $X(3872)$ is quite hot.
Due to the ambiguity information about the quantum numbers of $X(3872)$ at that time, someone held the view that
traditional quark model can still described its properties~\cite{05Barnes:2004,09Tornqvist:2004,07Swanson:2004,13Suzuki:2005}.
T. Barnes {\em et al.} analyzed some states by calculating their radiative transition, and the results turned out
that five states, $1^3D_3$, $1^3D_2$, $1^1D_2$, $2^3P_1$ and $2^1P_1$, could be the possible candidate of the new
exotic state~\cite{05Barnes:2004}. After examining the pictures of mesons and meson-meson molecules through different
ways, Pakasa and Swanson {\em et al.} concluded the meson picture may be more suitable for
$X(3872)$~\cite{06Pakasa:2004,07Swanson:2004}. Indeed, the ratio
$\frac{B(X(3872)\rightarrow\psi(2s)\gamma)}{B(X(3872)\rightarrow J/\psi\gamma)}=3.4\pm1.4$
announced by BABAR collaboration appears to support $X(3872)$ as a traditional meson~\cite{08Aubert:2009}.
N. Achasov argued that the $X(3872)$ resonance was the $c\bar{c}=\chi_{c1}(2P)$ charmonium which ``sit on" the
$D^{*0}\bar{D}^0$ threshold, and its mass was shifted from the prediction of potential model to the threshold of
$D^*\bar{D}$ by the contribution of the virtual $D^*\bar{D}+c.c.$ intermediate states~\cite{Achasov}.

Due to the mass of $X(3872)$ closes to the threshold of $D\bar{D}^*$, it's natural to classify state $X(3872)$
into molecule picture~\cite{09Tornqvist:2004}. T\"{o}rnqvist put forward that, analogous to deuteron,
one pion exchange potential could make contribution to bound state of $X(3872)$~\cite{09Tornqvist:2004},
and $X(3872)$ had structure of $\frac{1}{\sqrt{2}}[D\bar{D}^*+\bar{D}D^*]$. This picture could easily
explain the isospin breaking branching ratio
$\frac{Br(J/\psi\pi^-\pi^+\pi^0)}{Br(J/\psi\pi^+\pi^-)}=1.0\pm0.4\pm0.3$~\cite{10Abe:2005}
due to the mass difference between neutral $D$ meson and charged $D$ meson. Based on pion exchange potential
proposed by T\"{o}rnqvist, Swanson {\em et al.} considered additional mixture of $J/\psi\rho$ and $J/\psi\omega$
to the state, and arrived a conclusion that although the effect of pion exchange might be responsible for bound
state, short range quark dynamics were present and assisted in binding the $X(3872)$ via mixing to hidden charm
vector, $J/\psi$ states~\cite{11Swanson:2004}. However, some people held the different views of points that pion
exchange was too weak to
bind the $X(3872)$~\cite{12Liu:2008,13Suzuki:2005,14Thomas:2008}. Because $D^*-D-\pi=0$, Suzuki thought there was
no long rang attraction between $D$ and $D^*$, which means $X(3872)$ was not a bound state~\cite{13Suzuki:2005}.
What's more, unless the number of coupling constant and cut off were all big, $X(3872)$ could not be a real molecule
depending only on pion exchange potential~\cite{12Liu:2008}. Tomas {\em et al.} added tensor term and factor on the
OPE (one pion exchange) model, and they finally got a bound state~\cite{14Thomas:2008}.

Actually, since BABAR collaboration reported the branching ratio of $X\rightarrow\gamma J/\psi$, people gradually
started to accept the concept $X(3872)$ might be an unquenched hadron state, a mixture of $c\bar{c}$ and $D\bar{D}^*$\cite{18Kalashnikova:2005,19Ferretti:2013,20Ortega:2010,32cardoso:2014}. Kang and Oller analyzed the
experimental data by a near-threshold parameterization method and found the $X(3872)$ compositeness coefficients
in $D^0\bar{D}^{*0}$ ranges from nearly 0 up to 1 in the different scenarios~\cite{Kang}.
In 2005, Kalashnikova first used $^3P_0$ model to study the mass spectrum of $c\bar{c}$ based on
nonrelativistic quark model, in which the simple harmonic oscillator (SHO) functions are used to describe the
wavefunctions of mesons, and got the mass of $X(3872)$ 3990 MeV~\cite{18Kalashnikova:2005}, only 0.3 MeV
over threshold. In the following years, the $q\bar{q}\rightarrow qq\bar{q}\bar{q}$ transition operator in the
unquenched quark model is often taken as $^3P_0$ operator, which was first proposed by L. Micu~\cite{15Micu:1969},
then Yaouanc {\em et al.}
applied the operator to calculate the strong decay widths of baryons and mesons~\cite{16Yaouanc:1973,17Yaouanc:1974}.
Santopinto {\em et al.} calculated the decay width, components and mass of $X(3872)$ by
relativistic quark model and $^3P_0$ model, where SHO functions were also
used~\cite{19Ferretti:2013,40santopinto:2009,41santopinto:2012,42santopinto:2012,43santopinto:2015,44santopinto:2015}.
What's more, Ortega {\em et al.} utilized the unquenched quark model to analyze the decay width and components
of $X(3872)$, and their wavefunctions for the relative motion between two mesons were obtained by solving
the resonating group method (RGM) equation~\cite{20Ortega:2010,37Rupp:2012,38PGO:2012,39salamanca:2012}.
In the most of the previous work, the SHO functions are used to describe the meson dynamics and the relative motion
between two mesons are described by plane wave functions. The systematic errors due to the approximations
are unpredictable for the bound state calculation, although they are not a bad approximation for decay width
calculation~\cite{29chen:2018}. In this work, we applied the Gaussian expansion method (GEM), which is a powerful
method for few-body system, to analyze $X(3872)$ in an unquenched quark model~\cite{29chen:2018}, where a modified
$^3P_0$ was employed.

The paper is organized as follows. In section II, the chiral quark model and the GEM for solving the $q\bar q$
and $q\bar{q}$-$q\bar{q}$ systems are presented. In Sec. III, we briefly introduce the modified $^3P_0$ model. The
numerical results are given in Sec. IV. The last section is devoted to the summary of the present work.

\section{Chiral quark model and GEM} \label{GEM and chiral quark model}


In the chiral quark model~\cite{21Vijande:2005}, the meson spectrum is obtained by solving the
Schr\"{o}dinger equation
\begin{equation}
H \Psi_{IM_I}^{JM_J} (1,2) =E^{IJ} \Psi_{IM_I}^{JM_J} (1,2).
\end{equation}
The wavefunction $\Psi_{IM_I}^{JM_J}$ of a meson with quantum numbers $I^GJ^{PC}$ can be written as
\begin{equation}
\Psi_{IM_I}^{JM_J}(1,2) =\sum_{\alpha} \left[ \psi_{l}(\mathbf{r})\chi_{s}(1,2)\right]^{JM_J}
\omega^c(1,2)\phi_{IM_I}(1,2),
\end{equation}
where $\alpha$ denotes the intermediate quantum numbers $l,s$ and possible flavor indexes
(for isospin $I=0$ states, flavor indexes take $u\bar{u}, d\bar{d}$ and $s\bar{s}$).
"[~]" means angular momentum coupling, $\chi_{sm_s}(1,2),\omega^c(1,2),\phi_{IM_I}(1,2)$ are spin, color and
flavor wavefunctions (with specific isospin $I$) of meson, $\psi_{lm}(\mathbf{r}),\mathbf{r}=\mathbf{r}_1-\mathbf{r}_2$,
is the orbital wavefunction.
In GEM, the orbital wave function is written as the product of radial one and spherical harmonics, and the
radial part of the wave function is expanded by a set of gaussians,
\begin{eqnarray} \label{radialpart}
\psi_{lm}(\mathbf{r}) & = & \sum_{n=1}^{n_{max}} c_{n}\psi^G_{nlm}(\mathbf{r}), \nonumber \\
\psi^G_{nlm}(\mathbf{r}) & = & N_{nl}r^{l} e^{-\nu_{n}r^2}Y_l^m(\hat{\mathbf{r}}).
\end{eqnarray}
Gaussian size parameters are taken as the following geometric progression numbers
\begin{equation}\label{gaussiansize}
\nu_{n}=\frac{1}{r^2_n}, ~~~~ r_n=r_1a^{n-1},~~~~
a=\left(\frac{r_{n_{max}}}{r_1}\right)^{\frac{1}{n_{max}-1}}.
\end{equation}
This enables the optimization of ranges employing a small number of gaussians.
So, the wavefunction takes the form
\begin{eqnarray}
\Psi_{IM_I}^{JM}(1,2) & = & \sum_{n\alpha} C^{IJ}_{n\alpha} \Phi_{IM_I,n\alpha}^{JM} \nonumber \\
 & = & \sum_{n\alpha} C^{IJ}_{n\alpha}
 \left[ \psi_{l}(\mathbf{r})\chi_{s} \right]^{JM} \omega^c \phi_{IM_I}.
\end{eqnarray}
Noting that the gaussians are not orthogonal, the Rayleigh-Ritz variational principle
for solving the Schr\"{o}dinger equation leads to a generalized eigenvalue problem
\begin{eqnarray}
& & \sum_{n^{\prime},\alpha^{\prime}}(H_{n\alpha,n^{\prime}\alpha^{\prime}}^{IJ}
-E^{IJ} N_{n\alpha,n^{\prime}\alpha^{\prime}}^{IJ}) C_{n^{\prime}\alpha^{\prime}}^{IJ} = 0, \\
& & H_{n\alpha,n^{\prime}\alpha^{\prime}}^{IJ} =
  \langle\Phi^{JM}_{IM_I,n\alpha}| H | \Phi^{JM}_{IM_I,n^{\prime}\alpha^{\prime}}\rangle ,\\
 & & N_{n\alpha,n^{\prime}\alpha^{\prime}}^{IJ}=
  \langle\Phi^{JM}_{IM_I,n\alpha}| \Phi^{JM}_{IM_I,n^{\prime}\alpha^{\prime}}\rangle.
\end{eqnarray}

Extended to $q\bar{q}$-$q\bar{q}$ system, the same Schr\"{o}dinger equation is employed to obtained the
energy of the system,
\begin{equation}
    H \Psi^{JM_J}_{IM_I}(1234)=E^{IJ} \Psi^{JM_J}_{IM_I}(1234),
\end{equation}
where $\Psi^{JM_J}_{IM_I}$ is the wave function of the four-quark state, which can be constructed as follows.
First, we write down the wave functions of two subclusters,
\begin{eqnarray}
    \Psi^{J_1 M_{J_1}}_{I_1M_{I_1}}(12) =\left[ \psi_{l_1}(\mathbf{r}_{12})\chi_{s_1}(12)\right]^{J_1 M_{J_1}}
     \omega^{c_1}(12)\phi_{I_1M_{I_1}}(12),  \nonumber \\
    \Psi^{J_2 M_{J_2}}_{I_2M_{I_2}}(34) =\left[ \psi_{l_2}(\mathbf{r}_{34})\chi_{s_2}(34)\right]^{J_2 M_{J_2}}
    \omega^{c_2}(34)\phi_{I_2M_{I_2}}(34), \nonumber
\end{eqnarray}
where $\chi_{s},\omega^c,\phi^{I}$ are spin, color and flavor wave functions of the quark-antiquark subcluster
(the quarks are numbered as 1, 3 and antiquarks 2, 4). The bracket [~] denotes the angular momentum coupling.
Then, the total wave function of the four-quark state is obtained as:
\begin{eqnarray}
  & & \hspace{-10mm} \Psi^{JM_J}_{IM_I}(1234) = {\cal A} \left[\left[\psi_{l_1}(\mathbf{r}_{12})\chi_{s_1}(12)\right]^{J_1}
  \right. \nonumber \\
  & & \left.
      \left[\psi_{l_2}(\mathbf{r}_{34})\chi_{s_2}(34)\right]^{J_2}\psi_{L_r}(\mathbf{r}_{1234})\right]^{JM_J}
   \nonumber \\
  &   &  \left[\omega^{c_1}(12)\omega^{c_2}(34)\right]^{[222]} \left[\phi_{I_1}(12)\phi_{I_2}(34)\right]_{IM_I},
\end{eqnarray}
where $\psi_{L_r}(\mathbf{r}_{1234})$ is the wave function for the relative motion between two clusters with
orbital angular momentum $L_r$. ${\cal A}$ is the antisymmetrization operator. If all quarks (antiquarks) are
taken as identical particles, we have
\begin{equation}
{\cal A}=\frac{1}{2}(1-P_{13}-P_{24}+P_{13}P_{24}).
\end{equation}
The radial part of the wave function is also expanded by gaussians as in Eq.~(\ref{radialpart}). Finally, the
infinitesimally-shifted Gaussian basis function (ISG) is employed for the orbital wave functions with non-zero
orbital angular momentum to simplify the calculation of the matrix elements~\cite{30Hiyama:2003}:
\begin{eqnarray}
 \psi^G_{nlm}(\mathbf{r})& =& N_{nl}r^{l}
 e^{-\nu_{n}r^2}Y_l^m(\hat{\mathbf{r}}) \nonumber \\
 & =&N_{nl}\lim_{\epsilon \rightarrow
 0}\frac{1}{\epsilon^{l}}\sum_{k}^{k_{max}}C_{lm,k}e^{-\nu_{n}(\mathbf{r}-\epsilon\mathbf{D}_{lm,k})^2}.
\end{eqnarray}

The Hamiltonian of the chiral quark model includes three parts, the rest masses of quarks, the
nonrelativistic kinetic energy and the potential energy. The potential energy is composed of color confinement,
one-gluon-exchange (OGE) and one Goldstone boson exchange. The detailed form for the four-quark states is shown
below~\cite{21Vijande:2005}
\begin{eqnarray}
H &=& \sum_{i=1}^4
m_i+\frac{p_{12}^2}{2\mu_{12}}+\frac{p_{34}^2}{2\mu_{34}}
  +\frac{p_{1234}^2}{2\mu_{1234}}  \nonumber \\
  & & +\sum_{i<j=1}^4 \left( V_{ij}^{G}+V_{ij}^{C}+\sum_{\chi=\pi,K,\eta} V_{ij}^{\chi}
   +V_{ij}^{\sigma}\right), \nonumber\\
 V_{ij}^{G}&=& \frac{\alpha_s}{4} \boldsymbol{\lambda}_i^c \cdot \boldsymbol{\lambda}_{j}^c
\left[\frac{1}{r_{ij}}-\frac{2\pi}{3m_im_j}\boldsymbol{\sigma}_i\cdot
\boldsymbol{\sigma}_j
  \delta(\boldsymbol{r}_{ij})\right], \nonumber \\
  & &  \delta{(\boldsymbol{r}_{ij})}=\frac{e^{-r_{ij}/r_0(\mu_{ij})}}{4\pi r_{ij}r_0^2(\mu_{ij})},
  \nonumber \\
V_{ij}^{C}&=& ( -a_c r_{ij}^2-\Delta ) \boldsymbol{\lambda}_i^c
\cdot
 \boldsymbol{\lambda}_j^c \nonumber \\
V_{ij}^{\pi}&=& \frac{g_{ch}^2}{4\pi}\frac{m_{\pi}^2}{12m_im_j}
  \frac{\Lambda_{\pi}^2}{\Lambda_{\pi}^2-m_{\pi}^2}m_\pi v_{ij}^{\pi}
  \sum_{a=1}^3 \lambda_i^a \lambda_j^a,  \nonumber \\
V_{ij}^{K}&=& \frac{g_{ch}^2}{4\pi}\frac{m_{K}^2}{12m_im_j}
  \frac{\Lambda_K^2}{\Lambda_K^2-m_{K}^2}m_K v_{ij}^{K}
  \sum_{a=4}^7 \lambda_i^a \lambda_j^a,  \nonumber \\
V_{ij}^{\eta} & = &
\frac{g_{ch}^2}{4\pi}\frac{m_{\eta}^2}{12m_im_j}
\frac{\Lambda_{\eta}^2}{\Lambda_{\eta}^2-m_{\eta}^2}m_{\eta}
v_{ij}^{\eta} \nonumber \\
 && \left[\lambda_i^8 \lambda_j^8 \cos\theta_P
 - \lambda_i^0 \lambda_j^0 \sin \theta_P \right],  \nonumber \\
V_{ij}^{\sigma}&=& -\frac{g_{ch}^2}{4\pi}
\frac{\Lambda_{\sigma}^2}{\Lambda_{\sigma}^2-m_{\sigma}^2}m_\sigma
\left[
 Y(m_\sigma r_{ij})-\frac{\Lambda_{\sigma}}{m_\sigma}Y(\Lambda_{\sigma} r_{ij})\right] \nonumber \\
 v_{ij}^{\chi} & = & \left[ Y(m_\chi r_{ij})-
\frac{\Lambda_{\chi}^3}{m_{\chi}^3}Y(\Lambda_{\chi} r_{ij})
\right]
\boldsymbol{\sigma}_i \cdot\boldsymbol{\sigma}_j, \nonumber \\
& & Y(x)  =   e^{-x}/x,
\end{eqnarray}
where $m_i$ is the mass of quarks and antiquarks, and $\mu_{ij}$ is their reduced mass, $r_0(\mu_{ij})
=\hat{r}_0/\mu_{ij}$, $\boldsymbol{\sigma}$ are the $SU(2)$ Pauli matrices,
$\boldsymbol{\lambda},~\boldsymbol{\lambda}^c$ are $SU(3)$ flavor, color Gell-Mann matrices, $g^2_{ch}/4\pi$ is the
chiral coupling constant, determined from $\pi$-nucleon coupling constant. $\alpha_s$ is the effective scale-dependent
running quark-gluon coupling constant~\cite{21Vijande:2005},
\begin{equation}
\alpha_s(\mu_{ij})=\frac{\alpha_0}{\ln\left[(\mu_{ij}^2+\mu_0^2)/\Lambda_0^2\right]}
\end{equation}
All the parameters are determined by fitting the meson spectrum, from light to heavy, taking into account
only a quark-antiquark component. They are shown in Table~\ref{modelparameters}.

\begin{table}[t]
\begin{center}
\caption{Quark Model Parameters.\label{modelparameters}}
\begin{tabular}{cccc}
\hline\noalign{\smallskip}
Quark masses   &$m_u=m_d$(MeV)     &313  \\
               &$m_s$(MeV)         &536  \\
               &$m_c$(MeV)         &1728 \\
               &$m_b$(MeV)         &5112 \\
\hline
Goldstone bosons   &$m_{\pi}(fm^{-1})$     &0.70  \\
                   &$m_{\sigma}(fm^{-1})$     &3.42  \\
                   &$m_{\eta}(fm^{-1})$     &2.77  \\
                   &$m_{K}(fm^{-1})$     &2.51  \\
                   &$\Lambda_{\pi}=\Lambda_{\sigma}(fm^{-1})$     &4.2  \\
                   &$\Lambda_{\eta}=\Lambda_{K}(fm^{-1})$     &5.2  \\
                   &$g_{ch}^2/(4\pi)$                &0.54  \\
                   &$\theta_p(^\circ)$                &-15 \\
\hline
Confinement             &$a_c$(MeV)     &101 \\
                   &$\Delta$(MeV)       &-78.3 \\
\hline
OGE                 & $\alpha_0$        &3.67 \\
                   &$\Lambda_0(fm^{-1})$ &0.033 \\
                  &$\mu_0$(MeV)    &36.976 \\
                   &$\hat{r}_0$(MeV)    &28.17 \\
\hline
\end{tabular}
\end{center}
\end{table}

\section{$^3P_0$ model} \label{T operator}
 \subsection{traditional $^3P_0$ opertaor}
  The $^3P_0$ model (quark pair creation model) was originally introduced by Micu \cite{15Micu:1969} and further developed by Le
  Yaouanc, Ackleh and Roberts \textit{et al}.~\cite{16Yaouanc:1973,17Yaouanc:1974,22ACKleh:1996,23Roberts:1992}. It can be applied to the OZI
  rule allowed two-body strong decays of a hadron~\cite{15Micu:1969,22ACKleh:1996,24Capstic:1986,25Capstic:1994,26page:1995}. The
  transition operator in the model is,
  \begin{eqnarray} \label{T0}
  &&T_1=-3~\gamma\sum_m\langle 1m1-m|00\rangle\int
  d\mathbf{p}_3d\mathbf{p}_4\delta^3(\mathbf{p}_3+\mathbf{p}_4)\nonumber\\
  &&~~~~\times{\cal{Y}}^m_1(\frac{\mathbf{p}_3-\mathbf{p}_4}{2})
  \chi^{34}_{1-m}\phi^{34}_0\omega^{34}_0b^\dagger_3(\mathbf{p}_3)d^\dagger_4(\mathbf{p}_4),
 \end{eqnarray}
 where $\gamma$ represents the probability of the quark-antiquark pair with momentum $\mathbf{p}_3$ and $\mathbf{p}_4$
 created from the vacuum. Because the intrinsic parity of the
 antiquark is negative, the created quark-antiquark pair must be in the state $^{2S+1}L_{J}={}^3P_0$.
 $\phi^{34}_{0}$ and $\omega^{34}_0$ are flavor and color singlet states, respectively (the quark and the
 antiquark in the original meson are indexed by 1 and 2). The S-matrix element for the
 process $A \rightarrow B + C$ is written as
 \begin{equation}
 \langle BC|T|A\rangle=\delta^3(\mathbf{P}_A-\mathbf{P}_B-\mathbf{P}_C){\cal{M}}^{M_{J_A}M_{J_B}M_{J_C}},
 \end{equation}
 where $\mathbf{P}_B$ and $\mathbf{P}_C$ are the momenta of B and C mesons in the final state, and
 satisfy $\mathbf{P}_A = \mathbf{P}_B + \mathbf{P}_C = 0$ in the center-of-mass frame of meson A.
 ${\cal{M}}^{M_{J_A}M_{J_B}M_{J_C}}$ is the helicity amplitude of the process $A \rightarrow B + C$,
 which can be obtained as
 \begin{widetext}
 \begin{eqnarray}
 {\cal{M}}^{M_{J_A}M_{J_B}M_{J_C}}(\mathbf{P})&=&\gamma\sqrt{8E_AE_BE_C}
 \sum_{\renewcommand{\arraystretch}{.5}\begin{array}[t]{l}
 \scriptstyle M_{L_A},M_{S_A},\\\scriptstyle M_{L_B},M_{S_B},\\
 \scriptstyle M_{L_C},M_{S_C},m
 \end{array}}\renewcommand{\arraystretch}{1}\!\!
 \langle L_AM_{L_A}S_AM_{S_A}|J_AM_{J_A}\rangle
 \langle L_BM_{L_B}S_BM_{S_B}|J_BM_{J_B}\rangle\nonumber\\
 &&\times\langle
 L_CM_{L_C}S_CM_{S_C}|J_CM_{J_C}\rangle\langle 1m1-m|00\rangle
 \langle\chi^{14}_{S_BM_{S_B}}\chi^{32}_{S_CM_{S_C}}|\chi^{12}_{S_AM_{S_A}}\chi^{34}_{1-m}\rangle
 \nonumber\\
 &&\times[\langle \phi^{14}_B\phi^{32}_C|\phi^{12}_A\phi^{34}_0\rangle
 \mathcal{I}^{M_{L_A},m}_{M_{L_B},M_{L_C}}(\mathbf{P},m_1,m_2,m_3)\nonumber\\
 &&+(-1)^{1+S_A+S_B+S_C}\langle\phi^{32}_B\phi^{14}_C|\phi^{12}_A\phi^{34}_0\rangle
 \mathcal{I}^{M_{L_A},m}_{M_{L_B},M_{L_C}}(-\mathbf{P},m_2,m_1,m_3)],
 \end{eqnarray}
 with the momentum space integral
 \begin{eqnarray} \label{space integral}
 \mathcal{I}^{M_{L_A},m}_{M_{L_B},M_{L_C}}(\mathbf{P},m_1,m_2,m_3)=\int
 d\mathbf{p}\,\mbox{}\psi^\ast_{n_BL_BM_{L_B}}
 ({\scriptstyle\frac{m_3}{m_1+m_3}}\mathbf{P}+\mathbf{p})\psi^\ast_{n_CL_CM_{L_C}}
 ({\scriptstyle\frac{m_3}{m_2+m_3}}\mathbf{P}+\mathbf{p})
 \psi_{n_AL_AM_{L_A}}
 (\mathbf{P}+\mathbf{p}){\cal{Y}}^m_1(\mathbf{p}),\label{space}
 \end{eqnarray}
 \end{widetext}
 where $\mathbf{P}=\mathbf{P}_B=-\mathbf{P}_C$, and $\mathbf{p}=\mathbf{p}_3$, $m_3$ is the mass of the
 created quark $q_3$. To analyze the results and to compare the theoretical results with experimental data,
 the partial wave amplitude ${\mathcal{M}}^{JL}(A\rightarrow BC)$ is often employed. It is related with
 the helicity amplitude by the Jacob-Wick formula~\cite{27Jacob},
 \begin{eqnarray}
 &&{\mathcal{M}}^{J L}(A\rightarrow BC) = \frac{\sqrt{2 L+1}}{2 J_A
 +1} \!\! \sum_{M_{J_B},M_{J_C}} \langle L 0 J M_{J_A}|J_A
 M_{J_A}\rangle \nonumber\\&&\times\langle J_B M_{J_B}
 J_C M_{J_C} | J M_{J_A} \rangle \mathcal{M}^{M_{J_A} M_{J_B}
 M_{J_C}}({\textbf{P}}). \label{MJL}
 \end{eqnarray}
 In evaluating  the momentum space integral, Eq.~(\ref{space}), we use the wave functions of
 mesons $A$, $B$, $C$  obtained in the mass spectrum calculation. Because the  wave functions are
 expanded by a series of gaussians, the integral can be evaluated analytically.

The parameter $\gamma$ is generally determined by an overall fitting of the strong decay width of
hadrons. In this way, one obtains $\gamma=6.95$ for $u\bar{u}$ and $d\bar{d}$ pair creation, and
$\gamma=6.95/\sqrt{3}$ for $s\bar{s}$ pair creation \cite{28PLB}.

\subsection{Modified $^3P_0$ opertaor}
The modified transition operator $T_2$ (in position space) was first proposed by Chen {\em et al.}
for dealing with the issue that the mass shift of light meson is too large if the traditional
transition operator $T_1$ is used~\cite{29chen:2018}.
\begin{eqnarray} \label{T2}
   T_2&=&-3\gamma\sum_{m}\langle 1m1-m|00\rangle\int
   d\mathbf{r_3}d\mathbf{r_4}(\frac{1}{2\pi})^{\frac{3}{2}}ir2^{-\frac{5}{2}}f^{-5}
   \nonumber \\
   &&Y_{1m}(\hat{\mathbf{r}})e^{-\frac{r^2}{4f^2}}e^{-\frac{R_{AV}^2}{f_0^2}}\chi_{1-m}^{34}\phi_{0}^{34}
   \omega_{0}^{34}b_3^{\dagger}(\mathbf{r_3})d_4^{\dagger}(\mathbf{r_4}).
\end{eqnarray}
Here, $\mathbf{R}_{AV}=\mathbf{R}_A-\mathbf{R}_V$ is the relative coordinate between the source particle
``A" and the created quark-antiquark pair in the vacuum with
\begin{eqnarray}
   \mathbf{R}_A & = & \frac{m_1\mathbf{r_1}+m_2\mathbf{r_2}}{m_1+m_2}; \nonumber \\
   \mathbf{R}_V & = & \frac{m_3\mathbf{r_3}+m_4\mathbf{r_4}}{m_3+m_4}=\frac{\mathbf{r_3}+\mathbf{r_4}}{2}.
   \nonumber
\end{eqnarray}
The convergence factor $e^{-r^2/(4f^2)}$ of the modified operator $T_2$ mainly considers the effect
of quark-antiquark energy created in the vacuum, that it's difficult to create the quark-antiquark with
high energy. The damp factor $e^{-{R_{AV}}^2/{{R_{0}}}^2}$ takes into the fact that the created
quark-antiquark pair should not be far away from the source particle. With some reasonable arguments, the parameters
$f$ and $R_0$ were fixed, and the parameter $\gamma$ was determined by fitting the decay width of
$\rho \rightarrow \pi\pi$,
\begin{eqnarray*}
  \gamma=32.2~~f=0.5 ~\mbox{fm}~~R_{0}=1.0~ \mbox{fm}.
\end{eqnarray*}

Based on the modified transition operator, $T_2$, Chen {\em et al.} calculated the mass shift of light meson and
adjusted some parameters of quark model to make unquenched light ground state meson masses in agreement with
experimental data.
\begin{eqnarray*}
  \alpha_{0}=3.85,~~\Delta=-58.5
\end{eqnarray*}

\section{Unquenched Quark Model}

The mass and the structure of meson in unquenched quark model are obtained by solving the Schr\"{o}dinger equation
\begin{eqnarray}
   H\Psi_{IM_I}^{JM_J}=E\Psi_{IM_I}^{JM_J} ,
\end{eqnarray}
where, $\Psi_{IM_I}^{JM_J}$ is the unquenched wave function of the system which contain two- and four-quark components.
It can be written as:
\begin{eqnarray}
 \Psi_{IM_I}^{JM_J}=c_2\Psi_{IM_I}^{JM_J}(2q)+\sum_{i=1}^{N}c_{4i}\Psi_{i,IM_I}^{JM_J}(4q) ,
\end{eqnarray}
where $\Psi(2q)$ and $\Psi(4q)$ are the wave functions with two- and four-quark components, respectively
(the simplified symbols are used to save space), and the $N$ is the total number of four-quark channels.

In the nonrelativistic quark model, the number of particles is conserved. So there is no rigorous way to write
down the hamiltonian of the unquenched quark model. Here we only give a prescription of the Hamiltonian $H$ as follows:
\begin{eqnarray}
  H=H_{2q}+H_{4q}+T_{24} ,
\end{eqnarray}
where $H_{2q}$ is stipulated to act on the wave function of quark-antiquark component, $\Psi(2q)$, and $H_{4q}$ only
acts on the wave function of four-quark component, $\Psi(4q)$. $T_{24}$ takes the form of the transition operator in the
$^3P_0$ model, Eq.~(\ref{T0}) or Eq.~(\ref{T2}), which undertakes the coupling of the two- and four-quark components.
So, in this way, the matrix elements of the Hamiltonian can be written as:
\begin{eqnarray}
\langle\Psi|H|\Psi\rangle
  & = & c_2^2\langle\Psi(2q)|H_{2q}|\Psi(2q)\rangle \nonumber \\
  & + & \sum_{i,j=1}^{N}c_{4i}^{*}c_{4j} \langle\Psi_i(4q)|H_{4q}|\Psi_j(4q) \rangle    \nonumber \\
  & + & \sum_{i=1}^{N} c_{4i}^*c_2\langle\Psi_i(4q)|T_{24}|\Psi(2q)\rangle \nonumber \\
  & + & \sum_{j=1}^{N}c_2^*c_{4j}\langle\Psi(2q)|T_{24}^{\dagger}|\Psi_j(4q)\rangle.
\end{eqnarray}
Then, we get a blocked matrix of Hamiltonian and overlap:
\begin{eqnarray}
  (H)=\left[\begin{array}{ccccc} \langle H_{2q} \rangle  & \langle H_{24} \rangle_{1} & \langle H_{24} \rangle_{2} &...& \langle H_{24} \rangle_{n} \\
  \langle H_{42} \rangle_{1} & \langle H_{4q} \rangle_{11} & \langle H_{4q} \rangle_{12}&...& \langle H_{4q} \rangle_{1n}\\
  \langle H_{42} \rangle_{2} & \langle H_{4q} \rangle_{21} & \langle H_{4q} \rangle_{22}&...& \langle H_{4q} \rangle_{2n}\\
  ...&...&...&...&...\\
  \langle H_{42} \rangle_{n} & \langle H_{4q} \rangle_{n1} & \langle H_{4q} \rangle_{n2}&...& \langle H_{4q} \rangle_{nn}\\
  \end{array}
  \right]
\end{eqnarray}
\begin{eqnarray}
  (N)=\left[\begin{array}{ccccc} \langle N_{2q} \rangle  & 0 & 0 &...& 0 \\
  0 & \langle N_{4q} \rangle_{11} & \langle N_{4q} \rangle_{12}&...& \langle N_{4q} \rangle_{1n}\\
  0 & \langle N_{4q} \rangle_{21} & \langle N_{4q} \rangle_{22}&...& \langle N_{4q} \rangle_{2n}\\
  ...&...&...&...&...\\
  0 & \langle N_{4q} \rangle_{n1} & \langle N_{4q} \rangle_{n2}&...& \langle N_{4q} \rangle_{nn}\\
  \end{array}
  \right]
\end{eqnarray}
Where $\langle H_{2q} \rangle$, $\langle H_{24} \rangle_{j}$ and $\langle H_{4q} \rangle_{ij}$ are
$\langle\Psi(2q)|H_{2q}|\Psi(2q)\rangle$, $\langle \Psi(2q)|T_{24}|\Psi_j(4q)\rangle$, and
$\langle\Psi_i(4q)|H_{4q}|\Psi_j(4q)\rangle$ respectively. The subscript labels the index of the
four-quark channel. By solving the following generalized eigen-equation,
\begin{eqnarray}
  \Bigg( (H)-E_n(N) \Bigg) \Bigg( C_n \Bigg) = 0. \label{geig}
\end{eqnarray}
we get the eigen-energy $E_n$ and the expansion coefficients $C_n$.

\section{Numerical Results and discussions}

In the present calculation, we focus on the charmonium state $\chi_{c1}(2P)$ and try to explain the well-known exotic
state $X(3872)$ in the unquenched quark model. To fix the parameters associated with charm quark, two charmonia,
$\eta_c$ and $J/\psi$ are also investigated in the unquenched quark model. For comparison, two transition operators,
traditional one and the modified one, are used to do the calculation.

\subsection{Accumulating approach}

Generally, the dimension of $H$ in unquenched quark model is very big and matrix construction process is very complex.
So most of unquenched quark models adopted an accumulating approach, that is to do two channel coupling calculation,
bare state $\Psi(2q)$ and one of four-state $\Psi_i(4q)$, to get the mass shift $\Delta m_i$ and the probability fraction of
four-quark component $P_i$, then the total mass shift $\Delta M_t$ and the total fraction $P_t$ of four-quark
component are obtained by accumulating $\Delta m_i$ and $P_i$.
\begin{eqnarray}
  \Delta M_{t} & = & \sum_{i=1}^{N}\Delta m_{i}, \\
  P_{t} & = & \sum_{i=1}^{N}P_{i}.
\end{eqnarray}
In this approach, the dimension of the hamiltonian matrix $H$ needs to be diagonalized is reduced greatly.
The approach works well if the cross matrix elements between different four-quark channels are small enough.
Unfortunately, this is not always true. If there is coupling between two four-quark channels, the accumulating
approach will introduce errors, especially for the case that the energy of the four-quark state is close to
the bare mass of the meson. The comparison of the results for states $\eta_c$ and $\chi_{c1}(2P)$
between accumulating approach and the full-channel diagonalization is shown in Table \ref{try2},
where the traditional transition operator $T_1$ is used.
There is no coupling between $D\bar{D}^*$ and $D_s\bar{D}_s^*$, so the mass shifts and the probability fractions of
four-quark components for $\eta_c$ and $\chi_{c1}(2P)$ from the accumulating approach are almost the same as
that of three-channel diagonalization. Whereas there is a coupling between $D\bar{D}^*$ and $D^*\bar{D}$,
then different approaches give different results. For $\eta_c$, the accumulating approach over-estimates the mass shift
about 10\%, and the probability fraction 30\%. For $\chi_{c1}(2P)$, the over-estimation are 30\% for mass shift and 160\%
for probability fraction, because $D\bar{D}^*$ is open-channel for state $\chi_{c1}(2P)$. So the multi-channel coupling calculation
is adopted if there are non-zero matrix elements between different four-quark channels in the present work.

\begin{table}[tp]
\centering
\caption{The mass shifts and probability fractions of four-quark components in accumulating approach and the three-channel
 coupling calculation. 'S' in the parentheses denotes the relative motion between two clusters is in $S$-wave.\label{try2}}
\begin{tabular}{ccccc}
  \hline
              &  \multicolumn{2}{c}{$\eta_{c}$} & \multicolumn{2}{c}{$\chi_{c1}(2P)$}  \\
  bare mass   &  \multicolumn{2}{c}{2986.28}    & \multicolumn{2}{c}{3889.62} \\ \hline
  meson-meson & $\Delta m_i$ (MeV)  & $P_i$             & $\Delta m_i$ (MeV) & $P_i$         \\ \hline
  $D\bar{D}^{*}(S)$ & $-189.75$ & 4.02\%        & $-73.65$    & 43.09\%       \\
  $D_s^{*}\bar{D}_s(S)$ & $-76.50$ & 1.75\%        & $-15.13$    & 1.26\%     \\
  Total             & $-266.25$ & 5.77\%        & $-88.78$   & 44.35\%        \\ \hline
  $D\bar{D}^{*}$+$D_s^{*}\bar{D}_s~(S)$ & $-266.71$ & 5.20\%    & $-83.39$   & 35.53\% \\ \hline
  $D\bar{D}^{*}(S)$ & $-189.75$ & 4.02\%        & $-73.65$    &43.09\%        \\
  $D^{*}\bar{D}(S)$ & $-189.75$ & 4.02\%        & $-73.65$    &43.09\%        \\
  Total             & $-379.50$ & 8.04\%        & $-147.30$   &86.18\%        \\ \hline
  $D\bar{D}^{*}$+$D^{*}\bar{D}~(S)$ & $-369.43$ & 6.4\%    & $-117.11$ &  33.38\% \\ \hline
\end{tabular}
\end{table}

The comparison between our results of $\eta_c$ and that of other calculations are shown in Table \ref{result3},
where the same transition operator $T_1$ is employed.
Clearly our calculation gives larger mass shifts than other work. The reason is that the different wavefunctions
are used in the different calculations. In most previous work \cite{18Kalashnikova:2005,19Ferretti:2013,31Barnes:2008},
the simple harmonic oscillator (SHO) wavefunctions
are chosen as the radial part of the orbital wavefunctions of mesons, the relative motion wavefunction between two
meson clusters is set to the plane wave function. The plane wave approximation is a good one in the decay calculation,
it should not be a good one for the bound state calculation. In most case SHO wave function can describe the ground meson
well, but different size parameters should be used for different mesons. To simplify the calculation, the same size parameter
is used for different mesons in the most previous work. In our calculation, all the wavefunctions are determined by
system dynamics. In this a self-consistent calculation is arrived.
\begin{table}[tb]
\begin{center}
\caption{The mass shift of $\eta_{c}(1S)$ in the different calculations. (unit: MeV)}\label{result3}
\begin{tabular}{ccccc} \hline
  meson-meson & ~Ref.\cite{18Kalashnikova:2005}~ & ~Ref.\cite{19Ferretti:2013}~ &  ~Ref.\cite{31Barnes:2008}~ & This work \\ \hline
  $D \bar{D}^*(D^*\bar{D})$ & $-59$ & $-34$ & $-114$ & $-189.75$  \\ \hline
  $D^* \bar{D}^*$ & $-55$ & $-31$ & $-105$ & $-354.48$  \\ \hline
  $D_{s} \bar{D}_{s}^*(D_{s}^*\bar{D}_{s})$ & $-26$ & $-8 $ & $-106$ & $-76.50$ \\ \hline
  $D_{s}^* \bar{D}_{s}^* $ & $-35$ & $-8$ & $-98$ & $-147.14$ \\ \hline
\end{tabular}
\end{center}
\end{table}

\subsection{Results}

Using the modified operator $T_2$, and keeping all the parameters unchanged, we calculate the mass shifts and probability fractions
of four-quark components of state $\chi_{c1}(2P)$. The results are listed in Table \ref{result4}. From the table, we can see that
the open channel $D\bar{D}^*$ ($S$- and $D$-wave of the relative motion between $D$ and $\bar{D}^*$ are all considered) makes
the largest contribution, and pushes the bare mass of $\chi_{c1}(2P)$ down 76.37 MeV. The $D$-wave $D^*\bar{D}^*$ also pushes
the bare mass down 34.15 MeV. The mass shifts from the states $D_s\bar{D}_s^*$ and $D^*_s\bar{D}_s^*$ are small. At last the
four-quark components make mass shifts $-125.37$ MeV to the state $\chi_{c1}(2P)$, and the unquenched mass of the state is 3862.80 MeV,
which is close to mass of $X(3872)$. So to identify $X(3872)$ as $\chi_{c1}(2P)$ charmonium state in the unquenched quark model
is possible.
\begin{table}[tb]
\begin{center}
\caption{The mass shifts and probability fraction of four-quark components of $\chi_{c1}(2P)$. '$S$' and '$D$' in the parentheses
denote the relative motion between two clusters is in $S$-wave and $D$-wave. (unit: MeV)}\label{result4}
    \begin{tabular}{ccc} \hline
     \multicolumn{3}{c}{ $\chi_{c1}(2P)$} \\ \hline
    states & ~ $\Delta m_i$ ~& $P_i$ \\ \hline
    $D\bar{D}^{*}+D^*\bar{D}~(S+D)$ &~~~~$-76.37$~~~~&27.76\%\cr
    $D_s\bar{D}_s^{*}+D^*_s\bar{D}_s (S+D)$ &$-8.84$&1.98\%\cr
    $D^{*}\bar{D}^{*}~(D)$ &$-34.15$&6.42\%\cr
    $D_{s}^{*}\bar{D}_{s}^{*}~(D)$ &$-5.61$&0.72\%\cr
    Total & $-125.37$&36.88\%\\ \hline
    Unquenched mass &3862.80\\ \hline
\end{tabular}
\end{center}
\end{table}

\begin{table}[t]
\begin{center}
\caption{Adjusted quark model parameters.\label{3872parameters}}
\begin{tabular}{ccccc}
  \hline
  parameter  & & ChQM  &  Ref.\cite{29chen:2018}  & This work \\ \hline
  Quark masses  & $m_c$ (MeV)  & 1728  & 1728  & 1710 \\ \hline
  Confinement   & $a_c$ (MeV fm$^{-2}$) & 101   &   101  & 105.3 \\
                & $\Delta$ (MeV) & $-78.3$  & $-58.3$  & $-57.4$ \\ \hline
  $\alpha_{s}$  & $\alpha_{qq}$  & 0.57 & 0.60  & 0.60 \\
                & $\alpha_{qs}$  & 0.54 & 0.56  & 0.56 \\
                & $\alpha_{qc}$  & 0.49 & 0.52  & 0.52 \\
                & $\alpha_{sc}$  & 0.44 & 0.46  & 0.46 \\
                & $\alpha_{cc}$  & 0.38 & 0.39  & 0.39 \\ \hline
\end{tabular}
\end{center}
\end{table}

\begin{table*}[th]
\begin{center}
\caption{The mass shifts and probability fractions of four-quark components in UQM. '$S$', '$P$' and '$D$' in the parentheses
denotes the relative motion between two clusters are in $S$-, $P$- and $D$-wave. \label{try}}
\begin{tabular}{ccccccc}
  \hline
  State &  \multicolumn{2}{c}{$\eta_{c}$} & \multicolumn{2}{c}{$J/\psi$} & \multicolumn{2}{c}{$\chi_{c1}(2P)$} \\ \hline
  Bare mass & \multicolumn{2}{c}{3047.0} & \multicolumn{2}{c}{3169.7} & \multicolumn{2}{c}{3986.1} \\ \hline
  ~~~~~Meson-meson state~~~~~  & ~~~~$\Delta m_i$~~~~ & ~~~~$P_i$~~~~ & ~~~~$\Delta m_i$~~~~ & ~~~~$P_i$~~~~ &
   ~~~~$\Delta m_i$~~~~ & ~~~~$P_i$~~~~ \\ \hline
  $D\bar{D}^{*}+D^*\bar{D}~(P)$ & $-26.3$ & 2.5\% & $-20.5$ & 2.2\%  & - & - \\
  $D_s\bar{D}_s^*+D_s^*\bar{D}_s~(P)$ & $-6.0$ & 0.5\% & $-4.6$ & 0.4\% & - & - \\
  $D^{*}\bar{D}^{*}~(P)$ & $-24.8$ & 2.2\% & - & - & - & - \\
  $D_s^{*}\bar{D}_s^*~(P)$ & $-5.9$ & 0.5\% & - & - & - & - \\
  $D\bar{D}+D^*\bar{D}^*~(P)$ & - & - & $-39.4$ & 4.0\% & - & -  \\
  $D_s\bar{D}_s+D^*_s\bar{D}_s^*~(P)$ & - & - & $-9.0$ & 0.8\% & - & - \\
  $D\bar{D}^{*}+D^*\bar{D}~(S+D)$ & - & - & - & - & $-68.2$ & 22.5\% \\
  $D_s\bar{D}_s^*+D_s^*\bar{D}_s~(S+D)$ & - & - & - & - & $-8.2$ & 1.7\% \\
  $D^*\bar{D}^*~(D)$ & - & - & - & - & $-32.6$ & 5.6\% \\
  $D_s^*\bar{D}_s^*~(D)$ & - & - & - & - & $-5.4$ & 0.7\% \\
  Total & $-63.0$ & 5.7\% & $-73.5$ & 7.4\% & $-114.4$ & 30.5\% \\ \hline
  Unquenched mass & 2984.0 & 94.3\% & 3096.2 & 92.6\% & 3871.7& 69.5\% \\ \hline
\end{tabular}
\end{center}
\end{table*}

If we wang to reproduce the mass of the state $X(3872)$ exactly, we can fine tune the model parameters related
to the charm quark, and keep the light meson sector unchanged. To justify the fine-tuning, the charmonia $\eta_c$
and $J/\psi$ are also calculated in the unquenched quark model. The adjusted parameters are listed in
Table \ref{3872parameters} with the original parameters and the parameters adjusted in Ref.~\cite{29chen:2018}.
From the table, we can see that the adjustment is small, less than 5\% except the energy shift $\Delta$.

The results with new parameters for charmonia, $\eta_c$, $J/\psi$ and $\chi_{c1}(2P)$ are shown in
Table \ref{try}. The masses of $\eta_c$ and $J/\psi$ are fitted by adjusting the parameters, so the
experimental data are reproduced well, The dominant component is $c\bar{c}$, over 90\%.
In this case the calculated unquenched mass of $\chi_{c1}(2P)$
is 3871.7 MeV, almost the experimental value of $X(3872)$. In our calculation, the state is the mixing one
of $c\bar{c}$ and four-quark components. The dominant component is still $c\bar{c}$, $\sim 70\%$,
the fraction of $D\bar{D}^*+D^*\bar{D}$ is around 22.5\%, $D^*\bar{D}^*$, 5.6\% and $D_s^{(*)}\bar{D}_s^{(*)}$,
2.4\%. The results are consistent with some previous work~\cite{Achasov,19Ferretti:2013,Kang}

\section{Summary}
To describe the ordinary meson and exotic meson in one framework, an unquenched quark model is developed. As
a preliminary work, only four-quark components are taken into account, and the four-quark components are limited to
meson-meson states. To related the valence part to the high Fock components, the transition operator is needed.
Here the transition operator of $^3P_0$ model with modification is employed. The modification consists of two parts.
One considers the fact that the creation probability will decrease when the energy of created quark-pair increases.
Another requires that the created quark cannot be far away from the valence quark-pair. To minimize the error from
the calculation, a powerful method dealing with few-body system, GEM is used to find all the wavefunctions needed.
The unquenched quark model has been applied to the light meson spectrum, a reasonable result was obtained.
All the mass shifts are around 15\% of the bare masses of the states. In this way, the success of valence quark model
in describing low-lying spectrum of meson is kept. The present work applies the model to the charmonium states,
trying to explain the exotic state $X(3872)$.

By keeping the model parameters related to light meson unchanged, and fine-tuning the parameters related to charm quark,
we get the unquenched mass of $\chi_{c1}(2P)$ very close to the experimental value of $X(3872)$. At the same time,
the masses of charmonia, $\eta_c$ and $J/\psi$ are reproduced well. In our UQM, the high Fock components of ground
state charmonium $\eta_c$ and $J/\psi$ are small, less than 10\%. Similar results have been obtained for the light
mesons~\cite{29chen:2018}. For the states which can strong decay to two mesons, the probability fractions of two-meson
continua will be large, for example, the $\pi\pi$ continua in $\rho$-meson. Here for $X(3872)$, similar results are
obtained. The fraction of two-meson continua is around 30\%. However, the dominant component of $X(3872)$ is still
$c\bar{c}$, 70\%.

From our calculation, the unquenched quark model is a promising phenomenological method to unify the description of
ordinary mesons and exotic mesons. Of course further improvements are still needed, the four-quark components may be
hidden-color states or diquark-antidiquark states. To give a realistic description of all mesons, adjustment of
model parameters is expected. These are all our further work.

\end{document}